\documentclass[graphicx]{pasj00}
\Received{$\langle$reception date$\rangle$}
\Accepted{$\langle$acception date$\rangle$}
\Published{$\langle$publication date$\rangle$}
\SetRunningHead{Tsujimoto et al.}{Eu in the Draco dSph}

\usepackage{times}

\def\ltsima{$\; \buildrel < \over \sim\;$}
\def\ltsim{\lower.5ex\hbox{\ltsima}}
\def\gtsima{$\; \buildrel > \over\sim \;$}
\def\gtsim{\lower.5ex\hbox{\gtsima}}
\def\ms{$M_{\odot}$ }

\begin{document}

\title{Chemical Feature of Eu abundance in the Draco dwarf spheroidal galaxy\thanks{Based on data collected with the Subaru Telescope, which is operated by National Astronomical Observatory of Japan}}
\author{Takuji Tsujimoto\thanks{E-mail: taku.tsujimoto@nao.ac.jp}, \altaffilmark{1} Miho N. Ishigaki, \altaffilmark{2} Toshikazu Shigeyama, \altaffilmark{3} and Wako Aoki\altaffilmark{1}%
}
\altaffiltext{1}{%
   National Astronomical Observatory of Japan, 2--21--1 Osawa,\\
   Mitaka-shi, Tokyo 181--8588}
\altaffiltext{2}{%
    Kavli Institute for the Physics and Mathematics of the Universe (WPI), \\
    University of Tokyo, Kashiwa, Chiba 277-8583}
\altaffiltext{3}{%
    Research Center for the Early Universe, Graduate School of Science, \\
    University of Tokyo, 7-3-1 Hongo, Bunkyo-ku, Tokyo 113-0033}   

\maketitle

\begin{abstract}
Chemical abundance of r-process elements in nearby dwarf spheroidal (dSph) galaxies is a powerful tool to probe the site of r-process since their small-mass scale can sort out individual events producing r-process elements.  A merger of binary neutron stars is a  promising candidate of this site. In faint, or less massive dSph galaxies such as the Draco, a few binary neutron star mergers are expected to have occurred at most over the whole past. We have measured chemical abundances including Eu and Ba of three red giants in the Draco dSph by Subaru/HDS observation. The Eu detection for one star with [Fe/H]=$-1.45$ confirms a broadly constant [Eu/H] of $\sim-1.3$ for stars with [Fe/H]\gtsim$-2$. This feature is shared by other dSphs with similar masses, i.e., the Sculptor and the Carina, and suggests that neutron star merger is the origin of r-process elements in terms of its rarity. In addition, two very metal-poor stars with [Fe/H]=$-2.12$ and $-2.51$ are found to exhibit very low Eu abundances such as [Eu/H]$<-2$ with an implication of a sudden increase of Eu abundance by more than 0.7 dex at [Fe/H] $\approx -2.2$ in the Draco dSph. In addition, the detection of Ba abundances for these stars suggests that the r-process enrichment initiated no later than the time when only a few \% of stars in the present-day Draco dSph was formed. Though an identification of the origin of an early Eu production inside the Draco dSph should be awaited until more abundance data of stars with [Fe/H]\ltsim$-2$ in the Draco as well as other faint dSphs become available, the implied early emergence of Eu production event might be reconciled with the presence of extremely metal-poor stars enriched by r-process elements in the Galactic halo.
\end{abstract}

\section{Introduction}

Neutron star (NS) mergers as the origin of r-process elements have been highlighted thanks to the recent updates from  both observational and theoretical studies. First of all, the detection of a bump of a near infrared light in the afterglow of a short-duration $\gamma$-ray burst has become compelling evidence for the r-process synthesis in a NS merger event \citep{Barnes_13, Tanvir_13, Berger_13,  Hotokezaka_13}. Another advancement of our knowledge is brought by the results of theoretical nucleosynthesis calculations; an alternative candidate for the r-process site, neutrino-driven winds of core-collapse supernovae (CCSNe), have been found to become proton-rich due to revised treatments of neutrino transfer and unable to synthesize heavier r-process elements (A$>$130) \citep{Thompson_01, Wanajo_13} while recent studies have shown that NS mergers synthesize and eject them \citep{Korobkin_12, Bauswein_13, Wanajo_14}. In addition, chemical feature of r-process elements (Eu and Ba) in the Galaxy can be reproduced by enrichment proceeded through NS mergers in a novel model \citep[hereafter TT14]{Tsujimoto_14}, which improves the mixing process of r-process elements from the previous ones \citep{Mathews_90, Argast_04} to reflect the kinematic features of ejecta of NS mergers.

\begin{table*}[t]
\begin{center}
\begin{tabular}{lccccccc}
\hline
Name & RA & DEC & $V$ & Date  & ExpTime (sec)&  S/N\footnotemark & $V_{\rm los}$ (km s$^{-1}$)\\
\hline
\multicolumn{8}{c}{Draco targets}\\
\hline
Irwin 19826 &17 20 52.97&$+57$ 55 57.9&17.23& 2014/08/16 & 12900 &52 & $-292.5\pm 0.7$\\ 
Irwin 20751 &17 22 13.63&$+57$ 53 06.5&17.39& 2014/08/18 & 16200 &48 & $-301.5\pm 0.5$\\
Irwin 21275 &17 19 41.84&$+57$ 52 19.2&16.87& 2014/08/17 & 15300 &66 & $-295.0\pm 0.4$\\
\hline
\multicolumn{8}{c}{Comparison halo stars}\\
\hline
BD+30 2611& 15 06 53.83 & $+30$ 00 36.94 & 9.13 &2014/08/18& 600 & 238 & $-281.7\pm 0.3$\\
HD 122956 & 14 05 13.02& $-14$ 51 25.45 & 7.25 &2014/08/16& 200 & 195& \ \ \ $165.3\pm 0.3$\\ 
BD+09 2870& 14 16 29.95 & $+08$ 27 52.94 & 9.70 &2014/08/17& 600 & 325 & $-119.7\pm 0.2$\\
\hline
\multicolumn{8}{l}{$^1$a signal-to-noise per pixel ($\sim 5$ pixels per resolution element) measured by taking the standard deviation in}\\
\multicolumn{8}{l}{\ \ the wavelength range 6649-6653{\AA}.}\\
\end{tabular}
\caption{Summary of the observation}
\label{tab:obssummary}
\end{center}
\end{table*}

TT14 have proposed a more direct assessment of r-process origin by distinguishing between CCSNe and NS mergers from observed chemical abundances in nearby dwarf spheroidal (dSph) galaxies. Since NS merger events are considered to be rare, galaxies with small masses such as faint dSph galaxies are unlikely to leave so many NSs after the deaths of massive stars as to realize their mergers. For instance, a galaxy with its stellar mass of $10^5$ \ms are expected to host $\sim$ 500 CCSNe in total assuming a canonical stellar initial mass function (IMF). Then, the estimated frequency of NS mergers of one per 1000-2000 CCSNe (TT14) $<$1 NS merger in this galaxy. TT14 examined the observed trend of [Eu/H] against [Fe/H] for three faint dSphs, that is, the Draco, the Carina, the Sculptor, and found a constant [Eu/H] of $\sim -1.3$ with no apparent increase in the Eu abundance over the metallicity range of $-2$\ltsim [Fe/H]\ltsim$-1$. This result suggests that the r-process production is associated with events much rarer than SNe, such as  NS mergers. On the other hand, bright (i.e., massive) dSphs such as the Fornax and the Sagittarius exhibit a clear increasing [Eu/H] trend with respect to [Fe/H] as in the Milky Way, as might be expected since NS mergers are not so rare in these bright galaxies.

One puzzling question on this dSph matter is why and how these dSph stars are so promptly enriched with r-process elements up to [Eu/H]$\sim -1.3$, while these stars claim that no r-process productions operate during a major star formation epoch with [Fe/H] \gtsim $-2$ (see \$ 4). The idea that the gas forming these galaxies was already enriched 
at a level of [Eu/H]$\sim-1.3$ should be discarded since very metal-poor stars exhibit lower Eu abundance as observed for the star Draco 119 showing [Eu/H]$<-2.55$ with [Fe/H]=$-2.95$ \citep{Fulbright_04}. A few stars with [Fe/H]$<-3.5$ in the Sculptor also exhibit [Eu/H]$<-2.2$ \citep{Tafelmeyer_10, Simon_15}. Such low abundances of r-process element is verified by the Ba abundance for very metal-poor ([Fe/H]$<-2$) dSph stars \citep{Roederer_13}. Therefore it is likely that some kind of r-process synthesis may operate at a very early stage of star formation ([Fe/H]$\ltsim-2$) inside these dSphs and lift Eu abundance up to [Eu/H]$\sim -1.3$. Irrespective of its production site, we need to answer how events of r-process synthesis that occurred at a very early stage could leave no trace in stars with [Fe/H]\gtsim $-2$ which constitutes a main body of dSphs.

We aim at making a thorough investigation of the trend of r-process elements imprinted in stars covering the whole metallicity range in faint dSphs by measuring stellar Eu abundance. Note that Eu is an ideal element to trace the r-process enrichment; 97\% of the solar Eu abundance was synthesized by r-process, while the fraction of r-process is only 15\% in the solar Ba abundance \citep{Burris_00}. In particular, we will focus on (i) confirming a constant [Eu/H] ratio of $\sim -1.3$ for [Fe/H]\gtsim$-2$ and (ii) understanding how the [Eu/H] ratio increases up to $\sim -1.3$ for [Fe/H]\ltsim$-2$,  with a sufficient number of stars in the Draco and the Sextans dSphs.

This first paper presents results of chemical abundances measurements for three red giants in the Draco dSph and discusses its theoretical implications for the origin and enrichment history of r-process elements.

\section{Observation}

Three target red giant stars belonging to the Draco dSph were selected from \citet{Kirby_11}. For one of our target stars, Irwin 21275, an upper limit of Eu abundance has been estimated to be [Eu/H]$<-1.8$ \citep{Cohen_09}, while no Eu measurements were available for the other two stars. Observations were made with High-Dispersion Spectrograph (HDS) mounted on the Subaru Telescope \citep{Noguchi_02}. We adopted one of the standard setups of HDS, gStdYd", which covers a wavelength range of $\sim 4000-6800$ {\AA}. We use a slit width of $\timeform{1."1}$ to achieve a spectral resolving power of $R\sim 30000$. The $2\times 2$ binning was applied for our spectra. Summary of our observations of the three Draco stars together with comparison halo stars is tabulated in Table  \ref{tab:obssummary}.

\begin{figure}
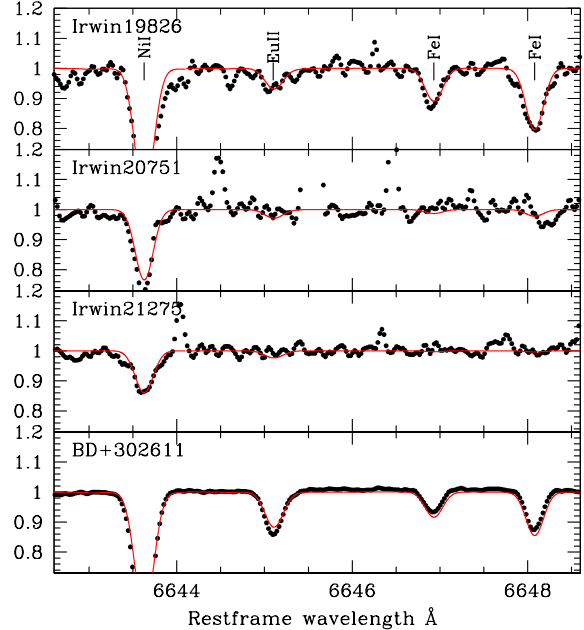

\begin{center}
\FigureFile( 85mm, 75mm ) {f1.eps}
\end{center}
\vspace{0.3cm}
\caption{Spectra of the Draco stars and one of the comparison stars (BD+30 2611) including the Eu II 6645 {\AA} absorption line. The red lines show synthetic spectra that give the best fits to the Eu II line for Irwin 19826 (top panel) and BD+30 2611 (bottom panel). For other two stars (two middle panels), synthetic spectra corresponding to a $2\sigma$ upper limit of the Eu abundance are shown.
}
\end{figure}

\begin{table*}
\begin{center}
\begin{tabular}{lcccccccc}
\hline
Name &  $T_{\rm eff}$ & $\log g$  &  $\xi$ & [Fe/H]\footnotemark[3] & $\log \epsilon$Eu & [Eu/H]\footnotemark[4] & $\log\epsilon$Ba & [Ba/H]\footnotemark[5] \\
& (K) & (dex) & (km s$^{-1}$) & (dex) &  (dex) & (dex)& (dex) & (dex)\\
\hline
Irwin 19826 &   4028 &  0.6 & 1.7 &  $-1.45\pm0.12$ & $-0.81$ &$-1.33\pm0.21$ & \ \ \ $0.04$& $-2.14\pm0.23$  \\
Irwin 20751 &  4278 & 0.6 & 1.7 & $-2.12\pm0.13$ & $<-1.5$ & $<-2.0$ & $-0.41$ & $-2.59\pm0.15$ \\
Irwin 21275 &   4406 &  0.7 & 2.3 & $-2.51\pm0.09$  & $<-1.6$ & $<-2.1$ &$-0.99$ & $-3.17\pm0.13$ \\
\hline
BD$+$30 2611  &   4201 & 0.6  &1.8& $-1.48\pm 0.12$&$-0.47$ &$-0.99\pm 0.14$ & $-$ & $-$\\
HD 122956   &   4595 & 1.3  & 1.7 & $-1.72\pm 0.12$& $-0.88$ & $-1.40\pm 0.14$ & $ \ \ \ 0.52$& $-1.66\pm 0.16$ \\
BD$+$09 2870  &   4612 & 1.2 & 2.0 & $-2.47\pm 0.09$& $<-2.45$($-2.64$) &$<-2.97$($-3.16$) & $-0.93$& $ -3.11\pm 0.11$ \\
\hline
\multicolumn{9}{l}{$^3$The weighted average of the estimates from neutral ([Fe I/H]) and ionized ([Fe II/H]) species. The uncertainty in each of}\\
\multicolumn{9}{l}{\ \ the [Fe I/H] and [Fe II/H] values is calculated as a quadratic sum of contributions from line-to-line scatter (root-mean}\\ 
\multicolumn{9}{l}{\ \ square difference from the mean value divided by a square-root of the number of lines) and uncertainties in atmospheric}\\ 
\multicolumn{9}{l}{\ \ parameters, which are assumed to be $\Delta T_{\rm eff}=100$ K, $\Delta \log g=0.3$ dex, and $\Delta \xi=0.2$ km s$^{-1}$.}\\
\multicolumn{9}{l}{$^4$The values in the parenthesis are deduced from the Eu II 4129.7 {\AA} line, while other values are from Eu II 6645.1 {\AA}.}\\
\multicolumn{9}{l}{\ \ Adopted solar abundance is $\log \epsilon$Eu$_{\odot}=0.52$ \citep{Asplund_09}. The uncertainty includes contributions from a}\\
\multicolumn{9}{l}{\ \ statistical error and uncertainties in the atmospheric parameters.}\\
\multicolumn{9}{l}{$^5$Adopted solar abundance is $\log\epsilon$Ba$_{\odot}=2.18$ \citep{Asplund_09}. The uncertainty is calculated by the same way as for}\\
\multicolumn{9}{l}{\ \ [Fe I/H].}
\end{tabular}
\caption{Atmospheric parameters and derived abundances}
\label{tab:stellarparams}
\end{center}
\end{table*}

Data were reduced with the HDS reduction pipeline ghdsql", which implements IRAF\footnote[2]{IRAF is distributed by the National Optical Astronomy Observatories, which are operated by the Association of Universities for Research in Astronomy, Inc. under cooperative agreement with the National Science Foundation.} scripts for cosmic-ray removal, flat fielding, scattered-light subtraction, and wavelength calibration. The reduced spectra are normalized by fitting the continuum fluxes with polynomial functions. Examples of the reduced spectra of the three program stars together with one comparison star are shown in Figure 1.

\section{Abundance analysis}

We performed a standard abundance analysis using a 1D-LTE analysis code adopted in \citet{Aoki_09}, which implements model atmospheres of \citet{Castelli_03}. The procedure is briefly summarized as follows (see Ishigaki et al. 2014 for details). First we measure equivalent widths of absorption lines by fitting a Gaussian profile to each absorption feature. The effective temperatures are estimated by $(V-K)$ color and an assumed Fe abundance, using the empirical formula for red giants \citep{Ramirez_05}. The $\log g$ values for the Draco stars are determined using the estimated $T_{\rm eff}$ by assuming that each star is located on a red giant branch of Yonsei-Yale isochrones \citep{Demarque_04} with an age 10.9 Gyr \citep{Orban_08}. Then we deduce Fe abundances from the equivalent widths of individual Fe I lines.  Here the micro-turbulence velocity parameter ($\xi$) is determined so as to meet the condition that Fe abundances dedudced from Fe I lines with different excitation energies do not depend on the equivalent widths. With these stellar parameters, we revise the effective temperature and iterate the same procedure until a consistent set of stellar parameters including the Fe abundance is obtained. The results are summarized in Table \ref{tab:stellarparams}. It should be of note that the estimated parameters for Irwin 21275 agree well with those estimated by \citet{Cohen_09}. The differences between the two analyses are $\Delta T_{\rm eff}=90$ K, $\Delta \log g=0.0$ dex, $\Delta \xi=0.1$ km s$^{-1}$, and $\Delta$ [Fe/H]$=0.1$ dex.

The Eu abundance is estimated by fitting synthetic spectra to each absorption feature. The wavelength region around the Eu II 4129.72 {\AA} line is too noisy for the Draco stars to define a continuum level. Instead we use the Eu II 6645.10 line for the Eu abundance estimate. The local continuum level is defined from the wavelength range of 6649.0-6653.0 {\AA}, in which strong absorption lines are expected to be absent for the stellar parameter range of our target stars. We inspected CN features in the spectra of the three stars, and found that their contribution is negligible in this wavelength range. The modification of the line profile due to the hyperfine structure is taken into account assuming an isotopic fraction for the two naturally existing isotopes, $^{151}$Eu and $^{153}$Eu, as $^{151}{\rm Eu}/(^{151}{\rm Eu}+^{153}{\rm Eu})=0.5$ \citep{Lawler_01}. Results of the fitting are shown in Figure 1. As shown in the two middle  panels, the Eu II lines are too weak to be detected for the two stars with [Fe/H]$<-2.0$ and thus only upper limits are deduced. On the other hand, for one metal-rich star with [Fe/H]$=-1.45$, the equivalent width due to the Eu II feature is estimated to be 20 m{\AA} while the $2 \sigma$ uncertainty estimated from the noise level of the continuum wavelength range is approximately 6 m{\AA}. Thus we argue that we detected the Eu II line and that its abundance is [Eu/H]$=-1.33$. The uncertainty in the equivalent width corresponds to an uncertainty in the Eu abundance of no more than $0.15$ dex. Changing each of the adopted atmospheric parameters by $\Delta T_{\rm eff}\pm 100$ K, $\Delta \log g\pm 0.3$ dex, and $\Delta \xi\pm 0.2$ km s$^{-1}$ results in $\Delta$[Eu/H] of 0.13 dex at most. The Ba abundances are estimated from the measured equivalent widths of Ba II 5853.69, 6141.73, 6496.91 {\AA} absorption lines. The effect of the hyperfine structure is also taken into account in the same way as in \citet{Ishigaki_13}. Results of the abundance measurements are summarized in Table \ref{tab:stellarparams}. Detailed elemental abundances including $\alpha$-, Fe-group, and other neutron-capture elements will be presented in our forthcoming paper (Ishigaki et al., in preparation).

\section{Eu feature and its implication}

\begin{figure}
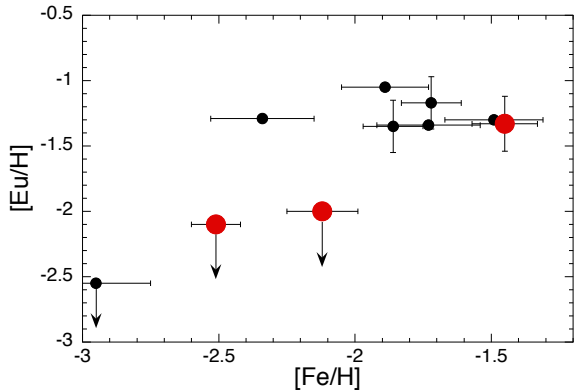

\vspace{0.2cm}
\begin{center}
\FigureFile( 75mm, 75mm ) {f2.eps}
\end{center}
\vspace{0.3cm}
\caption{Observed Eu abundances against Fe abundances in the Draco dSph. Our results are indicted by large filled circles, together with previous results by \citet{Shetrone_01}, \citet{Fulbright_04}, and \citet{Cohen_09}. 
}
\end{figure}

In this section, we present the chemical  feature of Eu abundance revealed by this study together with previous observations for not only the Draco dSph but also other two dSphs, i.e., the Sculptor and the Carina. Then, we try to advance our knowledge on the enrichment history of r-process elements as well as their origin. First we show our results for the Draco dSph in Figure 2. The Eu feature can be summarized as (i) the broadly constant Eu abundance of [Eu/H]$\sim-1.3$ for [Fe/H] \gtsim $-2$ and (ii) low Eu abundances of [Eu/H]\ltsim$-2$ deviating from a constant value for [Fe/H] \ltsim $-2.3 - -2.1$. A majority of  stars in the Draco dSph show the first feature, i.e, the constant Eu abundance, because the stellar metallicity distribution function \citep{Kirby_11} indicates that $\sim 85$\% of stars  populate the metallicity range of [Fe/H]$>-2.3$ or $\sim 70$\% of stars within [Fe/H]$>-2.1$. Thus the constant Eu abundance for [Fe/H] \gtsim $-2$ suggests a rarity of r-process production events. Using the Kroupa IMF \citep{Kroupa_93}, we obtain the corresponding stellar mass of $\sim 2.5\times 10^5$\ms \citep{Martin_08} and find that this stellar population presumably hosted $\sim$1400 CCSNe in total. Accordingly we conclude that the r-process production did not happen during sequential events of $\sim$1400 CCSNe in the Draco dSph. This implied rate is in good agreement with that of NS mergers, one per $1000-2000$ CCSNe, which is deduced from an analysis of chemical abundances in massive dSph galaxies (the Fornax and the Sagittarius) as well as in the Milky Way (TT14).

\begin{figure}
\vspace{0.2cm}
\begin{center}
\FigureFile( 75mm, 75mm ) {f3.eps}
\end{center}
\vspace{0.3cm}
\caption{observed Eu feature for the Draco (red circles), Sculptor (blue circles), and Carina (crosses) dSphs. The results acquired by this study are indicted by big red circles. Observed data are taken from \citet{Shetrone_03, Geisler_05, Kirby_12, Simon_15} for the Sculptor dSph, and \citet{Shetrone_03, Lemasle_12, Venn_12} for the Carina dSph.
}
\end{figure}

This broadly constant Eu feature breaks at a very low-metallicity range. Our result for Irwin 20751 indicates a very low [Eu/H] by more than $\sim 0.7$ dex at [Fe/H]=$-2.1\pm0.13$ as compared with the plateau abundance of [Eu/H]$\sim-1.3$. Since the star with [Fe/H]=$-2.34\pm0.19$ \citep{Cohen_09} already belongs to the plateau-like Eu abundance branch, [Eu/H] is likely to have an abrupt change at [Fe/H]$\approx-2.2$. Other two stars with [Fe/H]$<-2.5$ exhibiting low Eu abundances clearly confirm the presence of another branch of [Eu/H] for [Fe/H] \ltsim $-2$  far deviating from a constant [Eu/H] of $-1.3$. Note that Ba is detected in the spectra of our two stars with [Fe/H]$<-2$ as well as the star with [Fe/H]=$-2.95$ \citep{Fulbright_04}. This implies that these stars contain some amount of Eu. Accordingly, from an overall feature of Eu abundance, we conclude that the r-process operates inside the Draco dSph only at an early star formation with its enrichment up to [Eu/H]$\sim -1.3$, while a majority of stars formed in the subsequent epoch did not yield r-process elements. Regarding the site of an early Eu production, the feature of a sudden increase in [Eu/H] by more than 0.7 dex associated with little increase in [Fe/H] occurring around [Fe/H]=$-2.2$ seems compatible with the enrichment by a NS merger.

A common feature is shared with the Sculptor and Carina dSphs (Fig.~3). Clearly we see a constant [Eu/H] for [Fe/H]\gtsim $-2$ for the three dSphs. For [Fe/H]\ltsim $-2$, the Sclptor dSph also exhibits the low Eu abundance deviating from a constant value by  more than 3 dex, though the feature of a gradual increase in [Eu/H] between [Fe/H]$\approx -2.5$ and $-2$ is different from that of the Draco dSph. In addition, we present the evolution of [Ba/Eu] for the three dSphs in Figure 4. We see that dSph stars exhibit a clear increasing trend of [Ba/Eu], which is considered to be a result of a gradual enrichment of Ba by s-process, though a few stars including our star with [Fe/H]=$-1.45$ deviate from this trend. Since the observed lower limits of [Ba/Eu] ratios for the two stars with [Fe/H]$<-3$ are consistent with a pure r-process Ba/Eu ratio of [Ba/Eu]=$-0.89$ \citep{Burris_00}, we may be able to guess the values of [Eu/H] from their Ba abundances. Assuming that these two stars with [Fe/H]=$-4.05$ and $-3.52$ hold the pure r-process Ba/Eu ratio, we deduce individual [Eu/H] abundances of $-4.57$ and $-3.14$, respectively. It may give the bottom level of [Eu/H] abundance in faint dSphs. 

\begin{figure}
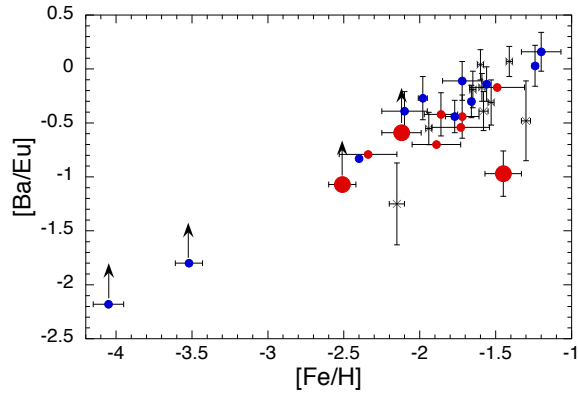

\vspace{0.2cm}
\begin{center}
\FigureFile( 75mm, 75mm ) {f4.eps}
\end{center}
\vspace{0.3cm}
\caption{Observed [Ba/Eu] ratios for the three dSphs. Symbols are the same as those in Fig.~3.
}
\end{figure}

Still too small data for [Fe/H]\ltsim$-2.5$ cast a veil over the evolutionary path of Eu abundance from the bottom, likely [Eu/H]$<-3$, to [Eu/H]$\sim -1.3$ and prevent us from identifying the site of the early Eu production inside faint dSphs. Even if we add information on the Ba feature (Fig.~5), its clear answer continues to elude us. It seems questionable to assume a selective early NS merger which should occur in each of the three dSphs, given its rarity. We may need to introduce some other contributors as an early r-process production such as  magneto-rotationally driven SNe \citep{Winteler_12, Wehmeyer_15}. To advance our understanding of the origin of early r-process production, more Eu data with precise measurements for stars with [Fe/H]\ltsim$-2$ in faint dSphs together with theoretical updates of the models for both NS merger and SNe are surely awaited. In particular, we can raise future prospect that improving the S/N values such as \gtsim 30 at the blue wave length of Eu II 4129.7 {\AA} likely achieved by 30m-class telescopes will reveal the Eu feature for very metal-poor regime. Putting aside an enigma on the site of an early Eu production, a clear signature of Eu enrichment imprinted on very early generations of dSph stars is reconciled with the fact that extremely-metal-poor halos stars in the Galaxy are already enriched in r-process, implying little time delay of r-process ejection as compared with other elements such as $\alpha$-elements or Fe.

\section{Concluding remarks}

From an assembly of Eu abundance data thus far obtained for three faint dSph galaxies including our new data for the Draco, it is clear that the trend of [Eu/H] against [Fe/H] shows completely different behaviors between two episodes, i.e., a very early phase and the subsequent major star formation period; a remarkable evolution of [Eu/H] from $<-4$ to $\sim-1.3$ in the metallicity range of [Fe/H]\ltsim$-2$ is followed by broadly constant [Eu/H] until the end of star formation. The latter feature exhibited by a main body of dSph stars that host more than 1000 CCSN  occurrences implies an extremely low event of the Eu production. Though NS merger is thought to satisfy this rarity as a source of r-process elements, there exist the Eu production events in the early phase of dSph formation, which  remains a mystery. Such an early commencement of r-process enrichment is likely to have something in common with the early r-process enrichment of the Galactic halo.

\begin{figure}
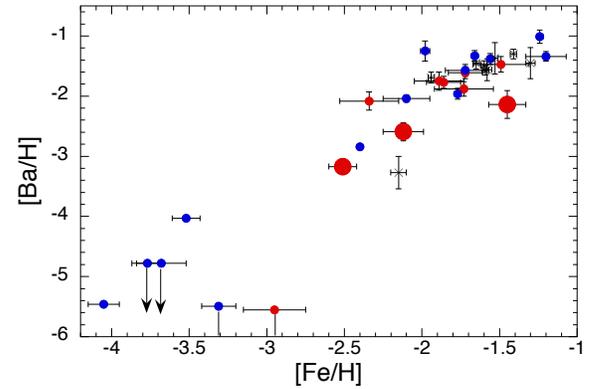

\vspace{0.2cm}
\begin{center}
\FigureFile( 75mm, 75mm ) {f5.eps}
\end{center}
\vspace{0.3cm}
\caption{Observed Ba feature for the three dSphs. Symbols are the same as those in Fig.~3.
}
\end{figure}

\bigskip
We are grateful to A. Tajitsu, K. Aoki, and staff members of the Subaru telescope for their helpful support and assistance in our HDS observation. This work was supported by JSPS KAKENHI Grant Number 23224004, 25-7047.

\end{document}